\documentclass[12pt]{article}%
\usepackage{amsmath}
\usepackage{amsfonts}
\usepackage{amssymb}
\usepackage{graphicx}
\usepackage[utf8]{inputenc}%
\providecommand{\U}[1]{\protect\rule{.1in}{.1in}}
\newtheorem{theorem}{Theorem}

\newtheorem{corollary}[theorem]{Corollary}

\newtheorem{lemma}[theorem]{Lemma}

\newtheorem{remark}[theorem]{Remark}

\newenvironment{proof}[1][Proof]{\noindent\textbf{#1.} }{\ \rule{0.5em}{0.5em}}
\begin{document}

\begin{titlepage}
\vspace{.3cm} \vspace{1cm}
\begin{center}
\baselineskip=16pt \centerline{\bf{\Large{ Geometry and the Quantum: Basics}}}\vspace{2truecm}
\centerline{\large\bf Ali H.
Chamseddine$^{1,3}$\ , Alain Connes$^{2,3,4}$ \ Viatcheslav Mukhanov$^{5,6}$\ \ } \vspace{.5truecm}
\emph{\centerline{$^{1}$Physics Department, American University of Beirut, Lebanon}}
\emph{\centerline{$^{2}$College de France, 3 rue Ulm, F75005, Paris, France}}
\emph{\centerline{$^{3}$I.H.E.S. F-91440 Bures-sur-Yvette, France}}
\emph{\centerline{$^{4}$Department of Mathematics, The Ohio State University, Columbus OH 43210 USA}}
\emph{\centerline{$^{5}$Theoretical Physics, Ludwig Maxmillians University,Theresienstr. 37, 80333 Munich, Germany }}
\emph{\centerline{$^{6}$MPI for Physics, Foehringer Ring, 6, 80850, Munich, Germany}}
\end{center}
\vspace{2cm}
\begin{center}
{\bf Abstract}
\end{center}
Motivated by the construction of spectral manifolds in noncommutative geometry, we introduce a higher degree Heisenberg commutation
relation involving the Dirac operator and the Feynman slash of scalar fields. This commutation relation appears in two versions, one sided and two sided. It implies  the quantization of the volume.  In the one-sided case it implies that the manifold
decomposes into a disconnected sum of spheres which will represent quanta of geometry. The two sided version in dimension $4$ predicts the two algebras $M_2(\mathbb H)$ and $M_4(\mathbb C)$ which are the algebraic constituents of the Standard Model of particle physics.  This taken together with the non-commutative algebra of functions allows one to reconstruct, using the spectral action, the Lagrangian of gravity coupled with the Standard Model. We show that any connected Riemannian Spin $4$-manifold with quantized volume $>4$ (in suitable units) appears as an irreducible representation of the two-sided commutation relations in dimension $4$ and that these representations give a seductive model of the ``particle picture" for a theory of quantum gravity in which both the Einstein geometric standpoint and the Standard Model emerge from Quantum Mechanics. Physical applications of this quantization scheme will follow in a separate publication.
\end{titlepage}
\tableofcontents
\section{Introduction}

The goal of this paper is to reconcile Quantum Mechanics and General
Relativity by showing that the latter naturally arises from a higher degree
version of the Heisenberg commutation relations. One great virtue of the
standard Hilbert space formalism of quantum mechanics is that it incorporates
in a natural manner the essential \textquotedblleft variability" which is the
characteristic feature of the Quantum: repeating twice the same experiment
will generally give different outcome, only the probability of such outcome is
predicted, the various possibilities form the spectrum of a self-adjoint
operator in Hilbert space. We have discovered a geometric analogue of the
Heisenberg commutation relations $[p,q]=i\hbar$. The role of the momentum $p$
is played by the Dirac operator. It takes the role of a measuring rod and at
an intuitive level it represents the inverse of the line element $ds$ familiar
in Riemannian geometry, in which only its square is specified in local
coordinates. In more physical terms this inverse is the propagator for
Euclidean Fermions and is akin to an infinitesimal as seen already in its
symbolic representation in Feynman diagrams where it appears as a solid (very)
short line $\,\;\;\bullet
\!\!\!-\!\!\!-\!\!\!-\!\!\!-\!\!\!-\!\!\!-\!\!\!\bullet\;\;$.

The role of the position variable $q$ was the most difficult to uncover. It
has been known for quite some time that in order to encode a geometric space
one can encode it by the algebra of functions (real or complex) acting in the
same Hilbert space as the above line element, in short one is dealing with
\textquotedblleft spectral triples". Spectral for obvious reasons and triples
because there are three ingredients: the algebra $\mathcal{A}$ of functions,
the Hilbert space $\mathcal{H}$ and the above Dirac operator $D$. It is easy
to explain why the algebra encodes a topological space. This follows because
the points of the space are just the characters of the algebra, evaluating a
function at a point $P\in X$ respects the algebraic operations of sum and
product of functions. The fact that one can measure distances between points
using the inverse line element $D$ is in the line of the Kantorovich duality
in the theory of optimal transport. It takes here a very simple form. Instead
of looking for the shortest path from point $P$ to point $P'$ as in Riemannian
Geometry, which only can treat path-wise connected spaces, one instead takes
the supremum of $|f(P)-f(P')|$ where the function $f$ is only constrained not
to vary too fast, and this is expressed by asking that the norm of the
commutator $[D,f]$ be $\leq1$. In the usual case where $D$ is the Dirac
operator the norm of $[D,f]$ is the supremum of the gradient of $f$ so that
the above control of the norm of the commutator $[D,f]$ means that $f$ is a
Lipschitz function with constant $1$, and one recovers the usual geodesic
distance. But a spectral triple has more information than just a topological
space and a metric, as can be already guessed from the need of a spin
structure to define the Dirac operator (due to Atiyah and Singer in that
context) on a Riemannian manifold. This additional information is the needed
extra choice involved in taking the square root of the Riemannian $ds^{2}$ in
the operator theoretic framework. The general theory is $K$-homology and it
naturally introduces decorations for a spectral triple such as a chirality
operator $\gamma$ in the case of even dimension and a charge conjugation
operator $J$ which is an antilinear isometry of $\mathcal{H}$ fulfilling
commutation relations with $D$ and $\gamma$ which depend upon the dimension
only modulo $8$. All this has been known for quite some time as well as the natural
occurrence of gravity coupled to matter using the spectral action applied to the 
tensor product  $\mathcal{A}\otimes A$ of the algebra $\mathcal{A}$ of functions by a 
finite dimensional algebra $A$ corresponding to internal structure.  In fact it was shown
in \cite{CC2} that one gets pretty close to zooming on the Standard Model of
particle physics when running through the list of irreducible spectral triples
for which the algebra $A$ is finite dimensional. The algebra that is
both conceptual and works for that purpose is
\[
A=M_{2}(\mathbb{H})\oplus M_{4}(\mathbb{C})
\]
where $\mathbb{H}$ is the algebra of quaternions and $M_{k}$ the matrices.
However it is fair to say that even if the above algebra is one of the first
in the list, it was not \emph{uniquely} singled out by our classification and
moreover presents the strange feature that the real dimensions of the two
pieces are not the same, it is $16$ for $M_{2}(\mathbb{H})$ and $32$ for
$M_{4}(\mathbb{C})$.

One of the byproducts of the present paper is a full understanding of this
strange choice, as we shall see shortly.

Now what should one beg for in a quest of reconciling gravity with quantum
mechanics? In our view such a reconciliation should not only produce gravity
but it should also naturally produce the other known forces, and they should
appear on the same footing as the gravitational force. This is asking a lot
and, in the minds of many, the incorporation of matter in the Lagrangian of
gravity has been seen as an unnecessary complication that can be postponed and
hidden under the rug for a while. As we shall now explain this is hiding the
message of the gauge sector which in its simplest algebraic understanding is
encoded by the above algebra $A=M_{2}(\mathbb{H})\oplus
M_{4}(\mathbb{C})$. The answer that we discovered is that the package formed
of the $4$-dimensional geometry together with the above algebra appears from a
very simple idea: to encode the analogue of the position variable $q$ in the
same way as the Dirac operator encodes the components of the momenta, just
using the Feynman slash. To be more precise we let $Y\in\mathcal{A}\otimes
C_{\kappa}$ be of the Feynman slashed form $Y=Y^{A}\Gamma_{A},$ and fulfill
the equations
\begin{equation}
Y^{2}=\kappa,\qquad Y^{\ast}=\kappa Y \label{zzz}%
\end{equation}
Here $\kappa=\pm1$ and $C_{\kappa}\subset M_{s}(\mathbb{C})$, $s=2^{n/2}$, is
the real algebra generated by $n+1$ gamma matrices $\Gamma_{A}$, $1\leq a\leq
n+1$\footnote{It is $n+1$ and not $n$ where $\Gamma_{n+1}$ is up to
normalization the product of the $n$ others.}
\[
\Gamma_{A}\in C_{\kappa},\quad\left\{  \Gamma^{A},\Gamma^{B}\right\}
=2\kappa\,\delta^{AB},\ (\Gamma^{A})^{\ast}=\kappa\Gamma^{A}%
\]
The one-sided higher analogue of the Heisenberg commutation relations is
\begin{equation}
\frac{1}{n!}\left\langle Y\left[  D,Y\right]  \cdots\left[  D,Y\right]
\right\rangle =\sqrt{\kappa}\,\gamma\quad\left(  n\mathrm{\ terms\,}\left[
D,Y\right]  \right)  \label{yyy}%
\end{equation}
where the notation $\left\langle T\right\rangle $ means the \emph{normalized}
trace of $T=T_{ij}$ with respect to the above matrix algebra $M_{s}%
(\mathbb{C})$ ($1/s$ times the sum of the $s$ diagonal terms $T_{ii}$). We
shall show below in Theorem \ref{bubles} that a solution of this equation
exists for the spectral triple $(\mathcal{A},\mathcal{H},D)$ associated to a
Spin compact Riemannian manifold $M$ (and with the components $Y^{A}%
\in\mathcal{A}$) if and only if the manifold $M$ breaks as the disjoint sum of
spheres of unit volume. This breaking into disjoint connected components
corresponds to the decomposition of the spectral triple into irreducible
components and we view these irreducible pieces as quanta of geometry. The
corresponding picture, with these disjoint quanta of Planck size is of course
quite remote from the standard geometry and the next step is to show that
connected geometries of arbitrarily large size are obtained by combining the
two different kinds of geometric quanta. This is done by refining the
one-sided equation \eqref{yyy} using the fundamental ingredient which is the
real structure of spectral triples, and is the mathematical incarnation of
charge conjugation in physics. It is encoded by an anti-unitary isometry $J$
of the Hilbert space $\mathcal{H}$ fulfilling suitable commutation relations
with $D$ and $\gamma$ and having the main property that it sends the algebra
$\mathcal{A}$ into its commutant as encoded by the \emph{order zero} condition
: $[a,JbJ^{-1}]=0$ for any $a,b\in\mathcal{A}$. This commutation relation
allows one to view the Hilbert space $\mathcal{H}$ as a bimodule over the
algebra $\mathcal{A}$ by making use of the additional representation $a\mapsto
Ja^{\ast}J^{-1}$. This leads to refine the quantization condition by taking
$J$ into account as the two-sided equation\footnote{The $\gamma$ involved here
commutes with the Clifford algebras and does not take into account an eventual
$\mathbb{Z}/2$-grading $\gamma_{F}$ of these algebras, yielding the full
grading $\gamma\otimes\gamma_{F}$.}
\begin{equation}
\frac{1}{n!}\left\langle Z\left[  D,Z\right]  \cdots\left[  D,Z\right]
\right\rangle =\gamma\quad Z=2EJEJ^{-1}-1,\ \label{jjj}%
\end{equation}
where $E$ is the spectral projection for $\{1,i\}\subset\mathbb{C}$ of the
double slash $Y=Y_{+}\oplus Y_{-}\in C^{\infty}(M,C_{+}\oplus C_{-})$. More
explicitly $E=\frac{1}{2}(1+Y_{+})\oplus\frac{1}{2}(1+iY_{-})$.

It is the classification of finite geometries of \cite{CC2} which suggested to
use the direct sum $C_{+}\oplus C_{-}$ of two Clifford algebras and the
algebra $C^{\infty}(M,C_{+}\oplus C_{-})$. As we shall show below in Theorem
\ref{bubles1} this condition still implies that the volume of $M$ is quantized
but no longer that $M$ breaks into small disjoint connected components. More
precisely let $M$ be a smooth connected oriented compact manifold of dimension
$n$. Let $\alpha$ be the volume form (of unit volume) of the sphere $S^{n}$.
One considers the (possibly empty) set $D(M)$ of pairs of smooth maps
$\phi_{\pm}:M\rightarrow S^{n}$ such that the differential form\footnote{We
use the notation $\phi^{\#}(\alpha)$ for the pullback of the differential form
$\alpha$ by the map $\phi$ rather than $\phi^{*}(\alpha)$ to avoid confusion
with the adjoint of operators.}
\[
\phi_{+}^{\#}(\alpha)+\phi_{-}^{\#}(\alpha)=\omega
\]
does not vanish anywhere on $M$ ($\omega(x)\neq0$ $\forall x\in M$). One
introduces an invariant $q(M)\subset\mathbb{Z}$ defined as the subset of
$\mathbb{Z}$:
\[
q(M):=\{\mathrm{degree}(\phi_{+})+\mathrm{degree}(\phi_{-})\mid(\phi_{+}%
,\phi_{-})\in D(M)\}\subset\mathbb{Z}.
\]
where $\mathrm{degree}(\phi)$ is the topological degree of the smooth map
$\phi$. Then a solution of \eqref{jjj} exists if and only if the volume of $M$
belongs to $q(M)\subset\mathbb{Z}$. We first check (Theorem \ref{threeman})
that $q(M)$ contains arbitrarily large numbers in the two relevant cases
$M=S^{4}$ and $M=N\times S^{1}$ where $N$ is an arbitrary connected compact
oriented smooth three manifold. We then give the proof (Theorem \ref{4man})
that the set $q(M)$ contains all integers $m\geq5$ for any smooth connected
compact spin $4$-manifold, which shows that our approach encodes all the
relevant geometries.

In the above formulation of the two-sided quantization equation the algebra
$C^{\infty}(M,C_{+}\oplus C_{-})$ appears as a byproduct of the use of the
Feynman slash. It is precisely at this point that the connection with our
previous work on the noncommutative geometry (NCG) understanding of the
Standard Model appears. Indeed as explained above we determined in \cite{CC2}
the algebra $A=M_{2}(\mathbb{H})\oplus M_{4}(\mathbb{C})$ as the
right one to obtain the Standard Model coupled to gravity from the spectral
action applied to the product space of a $4$-manifold $M$ by the finite space
encoded by the algebra $A$. Thus the full algebra is the algebra
$C^{\infty}(M,A)$ of $A$-valued functions on $M$. Now the
remarkable fact is that in dimension $4$ one has
\begin{equation}
C_{+}=M_{2}(\mathbb{H}),\ \ \ C_{-}=M_{4}(\mathbb{C}) \label{2algeb}%
\end{equation}
More precisely, the Clifford algebra $\mathrm{Cliff}(+,+,+,+,+)$ is the direct
sum of two copies of $M_{2}(\mathbb{H})$ and thus in an irreducible
representation, only one copy of $M_{2}(\mathbb{H})$ survives and gives the
algebra over $\mathbb{R}$ generated by the gamma matrices $\Gamma^{A}$. The
Clifford algebra $\mathrm{Cliff}(-,-,-,-,-)$ is $M_{4}(\mathbb{C})$ and it
also admits two irreducible representations (acting in a complex Hilbert
space) according to the linearity or anti-linearity of the way $\mathbb{C}$ is
acting. In both the algebra over $\mathbb{R}$ generated by the gamma matrices
$\Gamma^{A}$ is $M_{4}(\mathbb{C})$.

This fact clearly indicates that one is on the right track and in fact
together with the above two-sided equation it unveils the following tentative
\textquotedblleft particle picture" of gravity coupled with matter, emerging
naturally from the quantum world. First we now forget completely about the
manifold $M$ that was used above and take as our framework a fixed Hilbert
space in which $C=C_{+}\oplus C_{-}$ acts, as well as the grading $\gamma$,
and the anti-unitary $J$ all fulfilling suitable algebraic relations. So far
there is no variability but the stage is set. Now one introduces two
\textquotedblleft variables" $D$ and $Y=Y_{+}\oplus Y_{-}$ both self-adjoint
operators in Hilbert space. One assumes simple algebraic relations such as the
commutation of $C$ and $JCJ^{-1}$, of $Y$ and $JYJ^{-1}$, the fact that
$Y_{\pm}=\sum Y_{A}^{\pm}\Gamma_{\pm}^{A}$ with the $Y_{A}$ commuting with
$C$, and that $Y^{2}=1_{+}\oplus(-1)_{-}$ and also that the commutator $[D,Y]$
is bounded and its square again commutes with both $C_{\pm}$ and the
components $Y^{A}$, etc... One also assumes that the eigenvalues of the
operator $D$ grow as in dimension $4$. One can then write the two-sided
quantization equation \eqref{jjj} and show that solutions of this equation
give an emergent geometry. The geometric space appears from the joint spectrum
of the components $Y_{A}^{\pm}$. This would a priori yield an $8$-dimensional
space but the control of the commutators with $D$ allows one to show that it
is in fact a subspace of dimension $4$ of the product of two $4$-spheres. The
fundamental fact that the leading term in the Weyl asymptotics of eigenvalues
is \emph{quantized} remains true in this generality due to already developed
mathematical results on the Hochschild class of the Chern character in
$K$-homology. Moreover the strong embedding theorem of Whitney shows that
there is no a-priori obstruction to view the (Euclidean) space-time manifold
as encoded in the $8$-dimensional product of two $4$-spheres. The action
functional only uses the spectrum of $D$, it is the spectral action which,
since its leading term is now quantized, will give gravity coupled to matter
from its infinitesimal variation.

\section{Geometric quanta and the one-sided equation}

We recall that given a smooth compact oriented spin manifold $M$, the
associated spectral triple $(\mathcal{A},\mathcal{H},D)$ is given by the
action in the Hilbert space $\mathcal{H}=L^{2}(M,S)$ of $L^{2}$-spinors of the
algebra $\mathcal{A}=C^{\infty}(M)$ of smooth functions on $M$, and the Dirac
operator $D$ which in local coordinates is of the form
\begin{equation}
D=\gamma^{\mu}\left(  \frac{\partial}{\partial x^{\mu}}+\omega_{\mu}\right)
\end{equation}
where $\gamma^{\mu}=e_{a}^{\mu}\gamma^{a}$ and $\omega_{\mu}$ is the spin-connection.

\subsection{One sided equation and spheres of unit volume}

\begin{theorem}
\label{bubles} Let $M$ be a spin Riemannian manifold of even dimension $n$ and
$(\mathcal{A},\mathcal{H},D)$ the associated spectral triple. Then a solution
of the one-sided equation \eqref{yyy} exists if and only if $M$ breaks as the
disjoint sum of spheres of unit volume. On each of these irreducible
components the unit volume condition is the only constraint on the Riemannian
metric which is otherwise arbitrary for each component.
\end{theorem}

\proof We can assume that $\kappa=1$ since the other case follows by
multiplication by $i=\sqrt{-1}$. Equation \eqref{zzz} shows that a solution
$Y$ of the above equations gives a map $Y:M\rightarrow S^{n}$ from the
manifold $M$ to the $n$-sphere. Given $n$ operators $T_{j}\in\mathcal{C}$ in
an algebra $\mathcal{C}$ the multiple commutator
\[
\lbrack T_{1},\ldots,T_{n}]:=\sum\epsilon(\sigma)T_{\sigma(1)}\cdots
T_{\sigma(n)}%
\]
(where $\sigma$ runs through all permutations of $\{1,\ldots,n\}$) is a
multilinear totally antisymmetric function of the $T_{j}\in\mathcal{C}$. In
particular, if the $T_{i}=a_{i}^{j}S_{j}$ are linear combinations of $n$
elements $S_{j}\in\mathcal{C}$ one gets
\begin{equation}
\lbrack T_{1},\ldots,T_{n}]=\mathrm{Det}(a_{i}^{j})[S_{1},\ldots
,S_{n}]\label{antisym}%
\end{equation}
Let us compute the left hand side of \eqref{yyy}. The normalized trace of the
product of $n+1$ Gamma matrices is the totally antisymmetric tensor
\[
\left\langle \Gamma_{A}\Gamma_{B}\cdots\Gamma_{L}\right\rangle =i^{n/2}%
\epsilon_{AB\ldots L},\ \ A,B,\ldots,L\in\{1,\ldots,n+1\}
\]
One has $\left[  D,Y\right]  =\gamma^{\mu}\frac{\partial Y^{A}}{\partial
x^{\mu}}\Gamma_{A}=\nabla Y^{A}\Gamma_{A}$ where we let $\nabla f$ be the
Clifford multiplication by the gradient of $f$. Thus one gets at any $x\in M$
the equality
\begin{equation}
\left\langle Y\left[  D,Y\right]  \cdots\left[  D,Y\right]  \right\rangle
=i^{n/2}\epsilon_{AB\ldots L}Y^{A}\nabla Y^{B}\cdots\nabla Y^{L}%
\end{equation}
For fixed $A$, and $x\in M$ the sum over the other indices
\[
\epsilon_{AB\ldots L}Y^{A}\nabla Y^{B}\cdots\nabla Y^{L}=(-1)^{A}Y^{A}[\nabla
Y^{1},\nabla Y^{2},\ldots,\nabla Y^{n+1}]
\]
where all other indices are $\neq A$. At $x\in M$ one has $\nabla Y^{j}%
=\gamma^{\mu}\partial_{\mu}Y^{j}$ and by \eqref{antisym} the multi-commutator
(with $\nabla Y^{A}$ missing) gives
\[
\lbrack\nabla Y^{1},\nabla Y^{2},\ldots,\nabla Y^{n+1}]=\epsilon^{\mu\nu
\ldots\lambda}\partial_{\mu}Y^{1}\cdots\partial_{\lambda}Y^{n+1}[\gamma
^{1},\ldots,\gamma^{n}]
\]
Since $\gamma^{\mu}=e_{a}^{\mu}\gamma_{a}$ and $i^{n/2}[\gamma_{1}%
,\ldots,\gamma_{n}]=n!\gamma$ one thus gets by \eqref{antisym},
\begin{equation}
\left\langle Y\left[  D,Y\right]  \cdots\left[  D,Y\right]  \right\rangle
=n!\gamma\mathrm{Det}(e_{a}^{\alpha})\,\omega\label{expectY}%
\end{equation}
where
\[
\omega=\epsilon_{AB\ldots L}Y^{A}\partial_{1}Y^{B}\cdots\partial_{n}Y^{L}%
\]
so that $\omega dx_{1}\wedge\cdots\wedge dx_{n}$ is the pullback $Y^{\#}%
(\rho)$ by the map $Y:M\rightarrow S^{n}$ of the rotation invariant volume
form $\rho$ on the unit sphere $S^{n}$ given by
\[
\rho=\frac{1}{n!}\epsilon_{AB\ldots L}Y^{A}dY^{B}\wedge\cdots\wedge dY^{L}%
\]
Thus, using the inverse vierbein, the one-sided equation \eqref{yyy} is
equivalent to
\begin{equation}
\det\left(  e_{\mu}^{a}\right)  dx_{1}\wedge\cdots\wedge dx_{n}=Y^{\#}%
(\rho)\ \ \label{quant}%
\end{equation}
This equation \eqref{quant} implies that the Jacobian of the map
$Y:M\rightarrow S^{n}$ cannot vanish anywhere, and hence that the map $Y$ is a
covering. Since the sphere $S^{n}$ is simply connected for $n>1$, this implies
that on each connected component $M_{j}\subset M$ the restriction of the map
$Y$ to $M_{j}$ is a diffeomorphism. Moreover equation \eqref{quant} shows that
the volume of each component $M_{j}$ is the same as the volume $\int_{S^{n}%
}\rho$ of the sphere. Conversely it was shown in \cite{Connes} that, for
$n=2,4$, each Riemannian metric on $S^{n}$ whose volume form is the same as
for the unit sphere gives a solution to the above equation. In fact the above
discussion gives a direct proof of this fact for all (even) $n$. Since all
volume forms with same total volume are diffeomorphic (\cite{Moser}) one gets
the required result. \endproof

The spectral triple $(\mathcal{A},\mathcal{H},D)$ is then the direct sum of
the irreducible spectral triples associated to the components. Moreover one
can reconstruct the original algebra $\mathcal{A}$ as the algebra generated by
the components $Y^{A}$ of $Y$ together with the commutant of the operators
$D,Y,\Gamma_{A}$. This implies that a posteriori one recovers the algebra
$\mathcal{A}$ just from the representation of the $D,Y,\Gamma_{A}$ in Hilbert
space. As mentioned above the operator theoretic equation \eqref{yyy} implies
the integrality of the volume when the latter is expressed from the growth of
the eigenvalues of the operator $D$. Theorem \ref{bubles} gives a concrete
realization of this quantization of the volume by interpreting the integer $k$
as the number of geometric quantas forming the Riemannian geometry $M$. Each
geometric quantum is a sphere of arbitrary shape and unit volume (in Planck units).

\subsection{The degree and the index formula}

In fact the proof of Theorem \ref{bubles} gives a statement valid for any $Y$
not necessarily fulfilling the one-sided equation \eqref{yyy}. We use the
non-commutative integral as the operator theoretic expression of the
integration against the volume form $\det\left(  e_{\mu}^{a}\right)
dx_{1}\wedge\cdots\wedge dx_{n}$ of the oriented Riemannian manifold $M$. The
factor $2^{n/2+1}$ on the right comes from the factor $2$ in $Y=2e-1$ and from
the normalization (by $2^{-n/2}$) of the trace. The ${\int\!\!\!\!\!\! -} $ is
taken in the Hilbert space of the canonical spectral triple of the Riemannian manifold.

\begin{lemma}
\label{bublecor} For any $Y=Y^{A}\Gamma_{A},$ such that $Y^{2}=1$, $Y^{\ast
}=Y$ one has
\begin{equation}
{\int\!\!\!\!\!\!-}\gamma\left\langle Y\left[  D,Y\right]  ^{n}\right\rangle
D^{-n}=2^{n/2+1}\mathrm{degree}(Y) \label{expectY1}%
\end{equation}

\end{lemma}

\proof This follows from \eqref{expectY} which implies that
\[
\gamma\left\langle Y\left[  D,Y\right]  \cdots\left[  D,Y\right]
\right\rangle \det\left(  e_{\mu}^{a}\right)  dx_{1}\wedge\cdots\wedge
dx_{n}=n!\,Y^{\#}(\rho)
\]
while for any scalar function $f$ on $M$ one has (see \cite{Connesbook},
Chapter IV,2,$\beta$, Proposition 5), with $\Omega_{n}=2 \pi^{n/2}%
/\Gamma(n/2)$ the volume of the unit sphere $S^{n-1}$,
\[
{\int\!\!\!\!\!\! -} f D^{-n}=\frac1n (2\pi)^{-n}2^{n/2}\Omega_{n}\int_{M} f
\sqrt{g}dx^{n}
\]
Thus the left hand side of \eqref{expectY1} gives
\[
{\int\!\!\!\!\!\! -} \gamma\left\langle Y\left[  D,Y\right]  ^{n}\right\rangle
D^{-n}=\frac1n (2\pi)^{-n}2^{n/2}\Omega_{n} n!\int_{M} Y^{\#}(\rho)
\]
One has
\[
\int_{M} Y^{\#}(\rho)=\mathrm{degree}(Y)\Omega_{n+1}
\]
and
\[
\frac1n (2\pi)^{-n}2^{n/2}\Omega_{n} n! \Omega_{n+1}=2^{n/2+1}.
\]
using the Legendre duplication formula $2^{2z-1}\Gamma(z)\Gamma(z+\frac
12)=\sqrt{\pi}\Gamma(2z)$. \endproof

\section{Quantization of volume and the real structure $J$}

We consider the two sided equation \eqref{jjj}. The action of the algebra
$C_{+}\oplus C_{-}$ in the Hilbert space $H$ splits $H$ as a direct sum
$H=H^{(+)}\oplus H^{(-)} $ of two subspaces corresponding to the range of the
projections $1 \oplus0\in C_{+}\oplus C_{-}$ and $0 \oplus1\in C_{+}\oplus
C_{-}$. The real structure $J$ interchanges these two subspaces. The algebra
$C_{+}$ acts in $H^{(+)}$ and the formula $x\mapsto Jx^{*}J^{-1}$ gives a
right action of $C_{-}$ in $H^{(+)}$. We let $Y^{\prime}=iJY_{-}J^{-1}$ acting
in $H^{(+)}$ and $\Gamma^{\prime}=iJ\Gamma_{-}J^{-1}$ for the gamma matrices
of $C_{-}$. This allows us to reduce to the following simplified situation
occurring in $H^{(+)}$. We take $M$ of dimension $n=2m$ and consider two sets
of gamma matrices $\Gamma_{A}$ and $\Gamma_{B}^{\prime}$ which commute with
each other. We consider two fields
\begin{equation}
Y=Y^{A}\Gamma_{A},\ Y^{\prime}=Y^{\prime B}\Gamma_{B}^{\prime}\qquad
A,B=1,2,\ldots,n+1
\end{equation}
The condition $Y^{2}=1=Y^{\prime2}$ implies
\begin{equation}
Y^{A}Y^{A}=1,\ \ Y^{\prime B}Y^{\prime B}=1
\end{equation}
Let $e=\frac{1}{2}\left(  Y+1\right)  ,$ $e^{\prime}=\frac{1}{2}\left(
Y^{\prime}+1\right)  ,$ $E=ee^{\prime}=\frac{1}{2}\left(  Z+1\right)  $ then
$Z=2ee^{\prime}-1$ and thus%
\begin{align}
Z  &  =\frac{1}{2}\left(  Y+1\right)  \left(  Y^{\prime}+1\right)  -1\\
Z^{2}  &  =4e^{2}e^{^{\prime}2}-4ee^{\prime}+1=1
\end{align}
This means that $Z^{2}=1$ and we can use it to write the quantization
condition in the form
\begin{equation}
\frac{1}{n!}\left\langle Z\left[  D,Z\right]  ^{n} \right\rangle
=\gamma\label{yy2}%
\end{equation}
where $\left\langle {}\right\rangle $ is the normalized trace relative to the
matrix algebra generated by all the gamma matrices $\Gamma_{A}$ and
$\Gamma_{B}^{\prime}$.

\subsection{The normalized traces}

More precisely we let $\mathrm{Mat}_{+}$ be the matrix algebra generated by
all the gamma matrices $\Gamma_{A}$ and $\mathrm{Mat}_{-}$ be the matrix
algebra generated by all the gamma matrices $\Gamma_{B}^{\prime}$. We define
$\left\langle T\right\rangle _{\pm}$ as above as the \emph{normalized} trace,
which is $2^{-m}$ times the trace relative to the algebras $\mathrm{Mat}_{\pm
}$ of an operator $T$ in $H$. It is best expressed as an integral of the form
\begin{equation}
\left\langle T\right\rangle _{\pm}=\int_{\mathrm{Spin}_{\pm}}gTg^{-1}\,\,dg
\label{spin}%
\end{equation}
where $\mathrm{Spin}_{\pm}\subset\mathrm{Mat}_{\pm}$ is the spin group and
$dg$ the Haar measure of total mass $1$.

\begin{lemma}
\label{condexp} The conditional expectations $\left\langle T\right\rangle
_{\pm}$ fulfill the following properties

\begin{enumerate}
\label{condexp}

\item $\left\langle ST U\right\rangle _{+}=S \left\langle T \right\rangle
_{+}U$ for any operators $S,U$ commuting with $\mathrm{Mat}_{+}$ (this holds
similarly exchanging $+$ and $-$)

\item $\left\langle T \right\rangle =\left\langle \left\langle T \right\rangle
_{+}\right\rangle _{-}=\left\langle \left\langle T \right\rangle
_{-}\right\rangle _{+}$ for any operator $T$.

\item $\left\langle S T \right\rangle =\left\langle S \right\rangle
_{+}\left\langle T \right\rangle _{-}$ for any operator $S$ commuting with
$\mathrm{Mat}_{-}$ and $T$ commuting with $\mathrm{Mat}_{+}$.

\item $\left\langle S T \right\rangle =\left\langle S \right\rangle
_{-}\left\langle T \right\rangle _{+}$ for any operator $S$ commuting with
$\mathrm{Mat}_{+}$ and $T$ commuting with $\mathrm{Mat}_{-}$.
\end{enumerate}
\end{lemma}

\proof 1) follows from \eqref{spin} since $gSTUg^{-1}=SgTg^{-1}U$ for $S,U$
commuting with $\mathrm{Mat}_{+}$ and $g\in\mathrm{Spin}_{+}$.

2) The representation of the product group $G=\mathrm{Spin}_{+}\times
\mathrm{Spin}_{-}$ given by $(g,g^{\prime})\mapsto gg^{\prime}\in
\mathrm{Mat}_{+}\mathrm{Mat}_{-}$ is irreducible, and thus parallel to
\eqref{spin} one has
\begin{equation}
\label{spin1}\left\langle T\right\rangle =\int_{G}gg^{\prime}T(gg^{\prime
-1}dgdg^{\prime}=\left\langle \left\langle T \right\rangle _{+}\right\rangle
_{-}=\left\langle \left\langle T \right\rangle _{-}\right\rangle _{+}%
\end{equation}
using the fact that any $g$ commutes with any $g^{\prime}$.

3) This follows from \eqref{spin1} since one has
\[
gg^{\prime}ST(gg^{\prime-1}=gg^{\prime}S(gg^{\prime-1}gg^{\prime}%
T(gg^{\prime-1}=gSg^{-1}g^{\prime}Tg^{\prime-1}
\]

4) The proof is the same as for 3). \endproof

\subsection{Case of dimension $2$}

This is the simplest case, one has:

\begin{lemma}
The condition \eqref{yy2} implies that the ($2$-dimensional) volume of $M$ is
quantized. If $M$ is a smooth connected compact oriented $2$-dimensional
manifold with quantized volume there exists a solution of \eqref{yy2}.
\end{lemma}

\proof We shall compute the left hand side of \eqref{yy2} and show that
\begin{equation}
\left\langle Z\left[  D,Z\right]  \left[  D,Z\right]  \right\rangle =\frac
{1}{2}\left\langle Y\left[  D,Y\right]  \left[  D,Y\right]  \right\rangle
+\frac{1}{2}\left\langle Y^{\prime}\left[  D,Y^{\prime}\right]  \left[
D,Y^{\prime}\right]  \right\rangle \label{sum}%
\end{equation}
Thus as above we see that \eqref{yy2} is equivalent to the quantization
condition
\begin{equation}
\det\left(  e_{\mu}^{a}\right)  =\frac{1}{2}\epsilon^{\mu\nu}\epsilon
_{ABC}Y^{A}\partial_{\mu}Y^{B}\partial_{\nu}Y^{C}+\frac{1}{2}\epsilon^{\mu\nu
}\epsilon_{ABC}Y^{\prime A}\partial_{\mu}Y^{\prime B}\partial_{\nu}Y^{\prime
C} \label{quant2}%
\end{equation}
which gives the volume of $M$ as the sum of the degrees of the two maps
$Y:M\rightarrow S^{2}$ and $Y^{\prime}:M\rightarrow S^{2}$. This shows that
the volume is quantized (up to normalization). Conversely let $M$ be a compact
oriented $2$-dimensional manifold with quantized volume. Choose two smooth
maps $Y:M\rightarrow S^{2}$ and $Y^{\prime}:M\rightarrow S^{2}$ such that when
you add the pull back of the oriented volume form $\omega$ of $S^{2}$ by $Y$
and $Y^{\prime}$ you get the volume form of $M$. This will be discussed in
great details in \S \ref{diffgeom}. However, it is simple in dimension $2$
mostly because, on a connected compact smooth manifold, all smooth
nowhere-vanishing differential forms of top degree with the same integral are
equivalent by a diffeomorphism (\cite{Moser}). This solves equation
\eqref{quant2}. It remains to show \eqref{sum}. We use the properties%
\[
\left[  D,e\right]  =\left[  D,e^{2}\right]  =e\left[  D,e\right]  +\left[
D,e\right]  e
\]
which can be written as
\begin{equation}
e\left[  D,e\right]  =\left[  D,e\right]  (1-e),\ \ \left[  D,e\right]
e=(1-e)\left[  D,e\right]  \label{proj1}%
\end{equation}
which imply
\begin{equation}
e\left[  D,e\right]  e=0,\ \ e\left[  D,e\right]  ^{2}=\left[  D,e\right]
^{2}e \label{useform}%
\end{equation}
Now with $Z=2ee^{\prime}-1$ as above, one has
\begin{equation}
\left[  D,Z\right]  =2\left[  D,ee^{\prime}\right]  =2\left[  D,e\right]
e^{\prime}+2e\left[  D,e^{\prime}\right]
\end{equation}
Now $[D,e]$ commutes with $e^{\prime}$ because any element of $\mathrm{Mat}%
_{+}$ (such as $\Gamma^{A}$) commutes with any element of $\mathrm{Mat}_{-}$
(such as $\Gamma^{\prime B}$) and for any scalar functions $f,g$ one has
$[[D,f],g]=0]$ so that $[D,Y^{A}]$ commutes with $Y^{\prime B}$. Similarly
$[D,e^{\prime}]$ commutes with $e$ (and $e$ and $e^{\prime}$ commute) one thus
gets
\begin{align}
\left[  D,Z\right]  ^{2}  &  =4\left(  \left[  D,e\right]  e^{\prime}+e\left[
D,e^{\prime}\right]  \right)  ^{2}\nonumber\label{rememb}\\
&  =4\left(  \left[  D,e\right]  ^{2}e^{\prime}+e\left[  D,e^{\prime}\right]
^{2}+\left[  D,e\right]  ee^{\prime}\left[  D,e^{\prime}\right]  +\left[
D,e^{\prime}\right]  ee^{\prime}\left[  D,e\right]  \right)
\end{align}
One has
\begin{align}
\frac{1}{4}Z\left[  D,Z\right]  ^{2}  &  =e^{\prime}\left(  2e-1\right)
\left[  D,e\right]  ^{2}+e\left(  2e^{\prime}-1\right)  \left[  D,e^{\prime
}\right]  ^{2}\nonumber\\
&  +\left(  2e-1\right)  \left[  D,e\right]  ee^{\prime}\left[  D,e^{\prime
}\right]  +\left(  2e^{\prime}-1\right)  \left[  D,e^{\prime}\right]
e^{\prime}e\left[  D,e\right]
\end{align}
Using 4) of Lemma \ref{condexp}, one has
\[
\left\langle e^{\prime}\left(  2e-1\right)  \left[  D,e\right]  ^{2}%
\right\rangle =\left\langle e^{\prime}\right\rangle _{-}\left\langle \left(
2e-1\right)  \left[  D,e\right]  ^{2}\right\rangle _{+}=\frac{1}%
{2}\left\langle \left(  2e-1\right)  \left[  D,e\right]  ^{2}\right\rangle
\]
since $\left\langle (e-\frac{1}{2})\right\rangle _{-}=\frac{1}{2}\left\langle
Y^{\prime}\right\rangle _{-}=0$. Similarly one has
\[
\left\langle e\left(  2e^{\prime}-1\right)  \left[  D,e^{\prime}\right]
^{2}\right\rangle =\frac{1}{2}\left\langle \left(  2e^{\prime}-1\right)
\left[  D,e^{\prime}\right]  ^{2}\right\rangle _{-}=\frac{1}{2}\left\langle
\left(  2e^{\prime}-1\right)  \left[  D,e^{\prime}\right]  ^{2}\right\rangle
\]
Moreover one has $\left\langle Y\left[  D,Y\right]  \right\rangle =0$. This
follows from the order one condition since one gets, using $Y^{A}Y^{A}=1$,
\[
\left\langle Y\left[  D,Y\right]  \right\rangle =Y^{A}\left[  D,Y^{A}\right]
=\frac{1}{2}\left(  Y^{A}\left[  D,Y^{A}\right]  +\left[  D,Y^{A}\right]
Y^{A}\right)  =0.
\]
It implies that $\left\langle e\left[  D,e\right]  \right\rangle =0$ since it
is automatic that $\left\langle \left[  D,Y\right]  \right\rangle =0$. We then
get
\[
\left\langle \left(  2e-1\right)  \left[  D,e\right]  ee^{\prime}\left[
D,e^{\prime}\right]  \right\rangle =\left\langle \left(  2e-1\right)  \left[
D,e\right]  e\right\rangle _{+}\left\langle e^{\prime}\left[  D,e^{\prime
}\right]  \right\rangle _{-}=0
\]
and similarly for the other term. Thus we have shown that \eqref{sum} holds.
\endproof

\subsection{The two sided equation in dimension $4$}

This calculation will now be done for the four dimensional case:

\begin{lemma}
\label{4dim} In the $4$-dimensional case one has
\[
\left\langle Z\left[  D,Z\right]  ^{4} \right\rangle =\frac{1}{2}\left\langle
Y\left[  D,Y\right]  ^{4} \right\rangle +\frac{1}{2}\left\langle Y^{\prime
}\left[  D,Y^{\prime}\right]  ^{4} \right\rangle .
\]
The condition \ref{yy2} implies that the ($4$-dimensional) volume of $M$ is quantized.
\end{lemma}

\proof
Now $\Gamma_{A}$ and $\Gamma_{A}^{\prime}$ will have $A=1,\cdots,5.$ We now
compute, using \eqref{rememb}
\[
\frac{1}{16}\left[  D,Z\right]  ^{4} =\left(  \left[  D,e\right]
^{2}e^{\prime}+e\left[  D,e^{\prime}\right]  ^{2}+\left[  D,e\right]
ee^{\prime}\left[  D,e^{\prime}\right]  +\left[  D,e^{\prime}\right]
ee^{\prime}\left[  D,e\right]  \right)  ^{2}
\]
using \eqref{useform} to show that the following $6$ terms give $0$,
\[
(1)\times(4)= \left[  D,e\right]  ^{2}e^{\prime}\left[  D,e^{\prime}\right]
ee^{\prime}\left[  D,e\right]  =0, \ \ \text{since}\ \ e^{\prime}\left[
D,e^{\prime}\right]  e^{\prime}=0,
\]
\[
(2)\times(3)=e\left[  D,e^{\prime}\right]  ^{2}\left[  D,e\right]  ee^{\prime
}\left[  D,e^{\prime}\right]  =0, \ \ \text{since}\ \ e \left[  D,e \right]  e
=0,
\]
\[
(3)\times(1)=\left[  D,e\right]  ee^{\prime}\left[  D,e^{\prime}\right]
\left[  D,e\right]  ^{2}e^{\prime}=0, \ \ \text{since}\ \ e^{\prime}\left[
D,e^{\prime}\right]  e^{\prime}=0,
\]
\[
(3)\times(3)=\left[  D,e\right]  ee^{\prime}\left[  D,e^{\prime}\right]
\left[  D,e\right]  ee^{\prime}\left[  D,e^{\prime}\right]  =0,
\ \ \text{since}\ \ e^{\prime}\left[  D,e^{\prime}\right]  e^{\prime}=0,
\]
\[
(4)\times(2)=\left[  D,e^{\prime}\right]  ee^{\prime}\left[  D,e\right]
e\left[  D,e^{\prime}\right]  ^{2}=0, \ \ \text{since}\ \ e \left[  D,e
\right]  e =0,
\]
\[
(4)\times(4)=\left[  D,e^{\prime}\right]  ee^{\prime}\left[  D,e\right]
\left[  D,e^{\prime}\right]  ee^{\prime}\left[  D,e\right]  =0,
\ \ \text{since}\ \ e^{\prime}\left[  D,e^{\prime}\right]  e^{\prime}=0.
\]
We thus get the remaining ten terms in the form
\begin{align}
\frac{1}{16}\left[  D,Z\right]  ^{4}  &  =\left(  \left[  D,e\right]
^{2}e^{\prime}+e\left[  D,e^{\prime}\right]  ^{2}+\left[  D,e\right]
ee^{\prime}\left[  D,e^{\prime}\right]  +\left[  D,e^{\prime}\right]
ee^{\prime}\left[  D,e\right]  \right)  ^{2}\nonumber\\
&  =\left[  D,e\right]  ^{4}e^{\prime} +\left[  D,e\right]  ^{2}ee^{\prime
}\left[  D,e^{\prime}\right]  ^{2}+\left[  D,e\right]  ^{3}ee^{\prime}\left[
D,e^{\prime}\right] \nonumber\\
&  +\left[  D,e^{\prime}\right]  ^{2}e^{\prime}e\left[  D,e\right]
^{2}+e\left[  D,e^{\prime}\right]  ^{4}+\left[  D,e^{\prime}\right]
^{3}ee^{\prime}\left[  D,e\right] \nonumber\\
&  + \left[  D,e\right]  ee^{\prime} \left[  D,e^{\prime}\right]  ^{3}
+\left[  D,e\right]  ee^{\prime}\left[  D,e^{\prime}\right]  ^{2} e^{\prime
}e\left[  D,e\right] \nonumber\\
&  + \left[  D,e^{\prime}\right]  ee^{\prime} \left[  D,e\right]  ^{3}
+\left[  D,e^{\prime}\right]  ee^{\prime}\left[  D,e\right]  ^{2} e^{\prime
}e\left[  D,e^{\prime}\right]
\end{align}
We multiply by $Z=2ee^{\prime}-1$ on the left and treat the various terms as
follows.
\[
Z\left[  D,e\right]  ^{4}e^{\prime}=e^{\prime}(2e-1)\left[  D,e\right]  ^{4}
\]
gives the contribution
\[
\left\langle Z\left[  D,e\right]  ^{4}e^{\prime}\right\rangle =\left\langle
e^{\prime}\right\rangle \left\langle Y\left[  D,e\right]  ^{4}\right\rangle
=\frac{1}{32} \left\langle Y\left[  D,Y\right]  ^{4}\right\rangle
\]
The other quartic term
\[
Ze\left[  D,e^{\prime}\right]  ^{4}=e(2e^{\prime}-1)\left[  D,e^{\prime
}\right]  ^{4}
\]
gives the contribution
\[
\left\langle Ze\left[  D,e^{\prime}\right]  ^{4}\right\rangle =\frac{1}{32}
\left\langle Y^{\prime}\left[  D,Y^{\prime}\right]  ^{4}\right\rangle
\]
For the cubic terms one has, using $e \left[  D,e\right]  ^{3}e=e \left[
D,e\right]  e \left[  D,e\right]  ^{2}=0$,
\[
Z\left[  D,e\right]  ^{3}ee^{\prime}\left[  D,e^{\prime}\right]  =-\left[
D,e\right]  ^{3}ee^{\prime}\left[  D,e^{\prime}\right]
\]
and it gives as above a vanishing contribution since $\left\langle e^{\prime
}\left[  D,e^{\prime}\right]  \right\rangle =0$ (and similarly for $e$).
Similarly one has
\[
Z\left[  D,e^{\prime}\right]  ^{3}ee^{\prime}\left[  D,e\right]  =-\left[
D,e^{\prime}\right]  ^{3}ee^{\prime}\left[  D,e\right]
\]
which gives a vanishing contribution, as well as
\[
Z\left[  D,e\right]  ee^{\prime} \left[  D,e^{\prime}\right]  ^{3}=-\left[
D,e\right]  ee^{\prime} \left[  D,e^{\prime}\right]  ^{3}
\]
and
\[
Z\left[  D,e^{\prime}\right]  ee^{\prime} \left[  D,e\right]  ^{3}=-\left[
D,e^{\prime}\right]  ee^{\prime} \left[  D,e\right]  ^{3}.
\]

We now take care of the remaining $4$ quadratic terms. They are
\begin{align}
&  \left[  D,e\right]  ^{2}ee^{\prime}\left[  D,e^{\prime}\right]
^{2}+\left[  D,e^{\prime}\right]  ^{2}e^{\prime}e\left[  D,e\right]
^{2}\nonumber\\
&  +\left[  D,e\right]  ee^{\prime}\left[  D,e^{\prime}\right]  ^{2}e^{\prime
}e\left[  D,e\right]  +\left[  D,e^{\prime}\right]  ee^{\prime}\left[
D,e\right]  ^{2}e^{\prime}e\left[  D,e^{\prime}\right] \nonumber
\end{align}
One has, using the commutation of $ee^{\prime}$ with $\left[  D,e\right]
^{2}$
\[
Z\left[  D,e\right]  ^{2}ee^{\prime}\left[  D,e^{\prime}\right]  ^{2}=\left[
D,e\right]  ^{2}ee^{\prime}\left[  D,e^{\prime}\right]  ^{2}%
\]
so that the contributions of the two terms of the first line are
\begin{equation}
\left\langle e\left[  D,e\right]  ^{2}\right\rangle \left\langle e^{\prime
}\left[  D,e^{\prime}\right]  ^{2}\right\rangle +\left\langle e^{\prime
}\left[  D,e^{\prime}\right]  ^{2}\right\rangle \left\langle e\left[
D,e\right]  ^{2}\right\rangle \label{firstline}%
\end{equation}
Now for the remaining terms one gets, using $e\left[  D,e\right]  e=0$
\[
Z\left[  D,e\right]  ee^{\prime}\left[  D,e^{\prime}\right]  ^{2}e^{\prime
}e\left[  D,e\right]  =-\left[  D,e\right]  ee^{\prime}\left[  D,e^{\prime
}\right]  ^{2}e^{\prime}e\left[  D,e\right]
\]
To compute the trace one uses the fact that $\left[  D,e\right]  e$ commutes
with $\mathrm{Mat}_{-}$ and property $1)$ of Lemma \ref{condexp} to get
\[
\left\langle Z\left[  D,e\right]  ee^{\prime}\left[  D,e^{\prime}\right]
^{2}e^{\prime}e\left[  D,e\right]  \right\rangle _{-}=-\left[  D,e\right]
e\left\langle e^{\prime}\left[  D,e^{\prime}\right]  ^{2}\right\rangle
_{-}e\left[  D,e\right]
\]
Next one has, using $\left\langle Y^{\prime}\left[  D,Y^{\prime}\right]
^{2}\right\rangle =0$ and $e^{\prime}=\frac{1}{2}(Y^{\prime}+1)$,
\begin{equation}
\left\langle e^{\prime}\left[  D,e^{\prime}\right]  ^{2}\right\rangle
_{-}=\frac{1}{8}\left\langle \left[  D,Y^{\prime}\right]  ^{2}\right\rangle
\label{cliffsq}%
\end{equation}
and this does not vanish but is a scalar function which is $\sum\left[
D,Y^{\prime A}\right]  ^{2}$ and commutes with the other terms so that one
gets after taking it across
\[
\left\langle Z\left[  D,e\right]  ee^{\prime}\left[  D,e^{\prime}\right]
^{2}e^{\prime}e\left[  D,e\right]  \right\rangle =-\left\langle \left[
D,e\right]  e\left[  D,e\right]  \right\rangle \left\langle e^{\prime}\left[
D,e^{\prime}\right]  ^{2}\right\rangle
\]
Next one has, using $\left\langle Y\left[  D,Y\right]  ^{2}\right\rangle =0$,
and $Y\left[  D,Y\right]  +\left[  D,Y\right]  Y=0$
\[
\left\langle \left[  D,e\right]  e\left[  D,e\right]  \right\rangle =\frac
{1}{8}\left\langle \left[  D,Y\right]  ^{2}\right\rangle =\left\langle
e\left[  D,e\right]  ^{2}\right\rangle
\]
which shows that
\[
\left\langle Z\left[  D,e\right]  ee^{\prime}\left[  D,e^{\prime}\right]
^{2}e^{\prime}e\left[  D,e\right]  \right\rangle =-\left\langle e\left[
D,e\right]  ^{2}\right\rangle \left\langle e^{\prime}\left[  D,e^{\prime
}\right]  ^{2}\right\rangle
\]
Note that to show that
\[
\left\langle \left[  D,e\right]  e\left[  D,e\right]  \right\rangle
=\left\langle e\left[  D,e\right]  ^{2}\right\rangle
\]
one can also use (by \eqref{proj1})
\[
\left[  D,e\right]  e\left[  D,e\right]  =(1-e)\left[  D,e\right]
^{2},\ \ \left\langle (2e-1)\left[  D,e\right]  ^{2}\right\rangle
=\left\langle Y\left[  D,e\right]  ^{2}\right\rangle =0
\]
Similarly one gets
\[
\left\langle Z\left[  D,e^{\prime}\right]  ee^{\prime}\left[  D,e\right]
^{2}e^{\prime}e\left[  D,e^{\prime}\right]  \right\rangle =-\left\langle
e^{\prime}\left[  D,e^{\prime}\right]  ^{2}\right\rangle \left\langle e\left[
D,e\right]  ^{2}\right\rangle
\]
Thus combining with \eqref{firstline}, we get that the total contribution of
the quadratic terms is $0$.

Finally the second statement of Lemma \ref{4dim} follows from Lemma
\ref{bublecor}.\endproof

\subsection{Algebraic relations}

\label{algrel}

It is important to make the list of the algebraic relations which have been
used and do not follow from the definition of $Y$ and $Y^{\prime}$. Note first
that for $Y=Y^{A}\Gamma_{A}$ with the hypothesis that the components $Y^{A}$
belong to the commutant of the algebra generated by the $\Gamma_{B}$, one has
\[
Y^{2}=\pm1\implies[Y^{A},Y^{B}]=0, \ \forall A,B.
\]
Indeed the matrices $\Gamma_{A}\Gamma_{B}$ for $A<B$, are linearly independent
and the coefficient of $\Gamma_{A}\Gamma_{B}$ in the square $Y^{2}$ is
$[Y^{A},Y^{B}]$ which has to vanish. The similar statement holds for
$Y^{\prime}$. Moreover the commutation rule $[Y,Y^{\prime}]=0$ implies (and is
equivalent to) the commutation of the components $[Y^{A},Y^{\prime B}]=0$,
$\forall A,B$. Thus the components $Y^{A},Y^{\prime B}$ commute pairwise and
generate a commutative involutive algebra $\mathcal{A}$ (since they are all
self-adjoint). This corresponds to the order zero condition in the commutative
case. We have also assumed the order one condition in the from $[[D,a],b]=0$
for any $a,b\in\mathcal{A}$. But in fact we also made use of the commutation
of the operator $\left\langle \left[  D,Y\right]  ^{2}\right\rangle $ with the
elements of $\mathcal{A}$ and the $[D,a]$ for $a\in\mathcal{A}$ (and similarly
for $\left\langle \left[  D,Y^{\prime}\right]  ^{2}\right\rangle $).

\subsection{The Quantization Theorem}

In the next theorem the algebraic relations between $Y_{\pm}$, $D$, $J$,
$C_{\pm}$, $\gamma$ are assumed to hold. We shall not detail these relations
but they are exactly those discussed in \S \ref{algrel} and which make the
proof of Lemma \ref{4dim} possible.

As in the introduction we adopt the following definitions. Let $M$ be a
connected smooth oriented compact manifold of dimension $n$. Let $\alpha$ be
the volume form of the sphere $S^{n}$. One considers the (possibly empty) set
$D(M)$ of pairs of smooth maps $\phi,\psi:M\to S^{n}$ such that the
differential form
\[
\phi^{\#}(\alpha)+\psi^{\#}(\alpha)=\omega
\]
does not vanish anywhere on $M$ ($\omega(x)\neq0$ $\forall x\in M$). One
defines an invariant which is the subset of $\mathbb{Z}$:
\[
q(M):=\{\mathrm{degree}(\phi)+\mathrm{degree}(\psi)\mid(\phi,\psi)\in
D(M)\}\subset\mathbb{Z}.
\]

\begin{theorem}
\label{bubles1} Let $n=2$ or $n=4$.

$(i)$~In any operator representation of the two sided equation \eqref{jjj} in
which the spectrum of $D$ grows as in dimension $n$ the volume (the leading
term of the Weyl asymptotic formula) is quantized.

$(ii)$~Let $M$ be a connected smooth compact oriented spin Riemannian manifold
(of dimension $n=2,4$). Then a solution of \eqref{jjj} exists if and only if
the volume of $M$ is quantized\thinspace\thinspace\footnote{up to
normalization} to belong to the invariant $q(M)\subset\mathbb{Z}$.
\end{theorem}

\proof $(i)$~By Lemma \ref{4dim} one has, as in the two dimensional case that
the left hand side of \eqref{yy2} is up to normalization,
\begin{equation}
L=\left\langle Y\left[  D,Y\right]  ^{4}\right\rangle +\left\langle Y^{\prime
}\left[  D,Y^{\prime}\right]  ^{4} \right\rangle \label{sum1}%
\end{equation}
so that \eqref{yy2} implies that the volume of $M$ is (up to sign) the sum of
the degrees of the two maps. This is enough to give the proof in the case of
the spectral triple of a manifold, and we shall see in Theorem \ref{thm3} that
it also holds in the abstract framework.

$(ii)$~Using Lemma \ref{4dim} the proof is the same as in the two dimensional
case. Note that the connectedness hypothesis is crucial in order to apply the
result of \cite{Moser}.\endproof

\section{Differential geometry and the two sided equation}

\label{diffgeom}

The invariant $q_{M}$ makes sense in any dimension. For $n=2,3$, and any
connected $M$, it contains all sufficiently large integers. The case $n=4$ is
more difficult but we shall prove below in Theorem \ref{4man} that it contains
all integers $m>4$ as soon as the connected $4$-manifold $M$ is a Spin
manifold, an hypothesis which is automatic in our context.

\subsection{Case of dimension $n<4$}

\begin{lemma}
\label{thm2} Let $M$ be a compact connected smooth oriented manifold of
dimension $n<4$. Then for any differential form $\omega\in\Omega^{n}(M)$ which
vanishes nowhere, agrees with the orientation, and fulfills the quantization
$\int_{M}\omega\in\mathbb{Z}$, $\vert\int_{M}\omega\vert>3$, one can find two
smooth maps $\phi,\phi^{\prime}$ such that
\[
\phi^{\#}(\alpha)+\phi^{\prime\#}(\alpha)=\omega
\]
where $\alpha$ is the volume form of the sphere of unit volume.
\end{lemma}

\proof By \cite{Alex} as refined in \cite{Ramirez}, any Whitehead
triangulation of $M$ provides (after a barycentric subdivision) a ramified
covering of the sphere $S^{n}$ obtained by gluing two copies $\Delta^{n}_{\pm
}$ of the standard simplex $\Delta^{n}$ along their boundary. One uses the
labeling of the vertices of each $n$-simplex by $\{0,1,\ldots,n\}$ where each
vertex is labeled by the dimension of the face of which it is the barycenter.
The bi-coloring corresponds to affecting each $n$-simplex of the triangulation
with a sign depending on wether the orientation of the simplex agrees or not
with the orientation given by the labeling of the vertices. One then gets a
PL-map $M\to S^{n}$ by mapping each simplex with a $\pm$ sign to $\Delta
^{n}_{\pm}$ respecting the labeling of the vertices. This gives a covering
which is ramified only on the $(n-2)$-skeleton of $\Delta^{n}_{\pm}$. After
smoothing one then gets a smooth map $\phi:M\to S^{n}$ whose Jacobian will be
$>0$ outside a subset $K$ of dimension $n-2$ of $M$. Using the hypothesis
$n<4$ (which gives $(n-2)+(n-2)<n$), the set of orientation preserving
diffeomorphisms $\psi\in\mathrm{Diff}^{+}(M)$ such that $\psi(K)\cap
K=\emptyset$ is a dense subset of $\mathrm{Diff}(M)^{+}$, thus one finds
$\psi\in\mathrm{Diff}^{+}(M)$ such that the Jacobian of $\phi$ and the
Jacobian of $\phi^{\prime}=\phi\circ\psi$ never vanish simultaneously. This
shows that the differential form $\rho=\phi^{\#}(\alpha)+\phi^{\prime
\#}(\alpha)$ does not vanish anywhere and by the result of \cite{Moser} there
exists an orientation preserving diffeomorphism of $M$ which transforms this
form into $\omega$ provided they have the same integral. But the integral of
$\rho$ is twice the integral of $\phi^{\#}(\alpha)$ which in turns is the
degree of the map $\phi$ and thus the number of simplices of a given color. As
performed the above construction only gives even numbers, since the integral
of $\rho$ is twice the degree of the map $\phi$, but we shall see shortly in
Lemma \ref{single} that in fact the degree of the map $\phi$ is in $q(M)$ from
a fairly general argument. \endproof

\begin{figure}[ptb]
%\begin{center}
\includegraphics[scale=1]{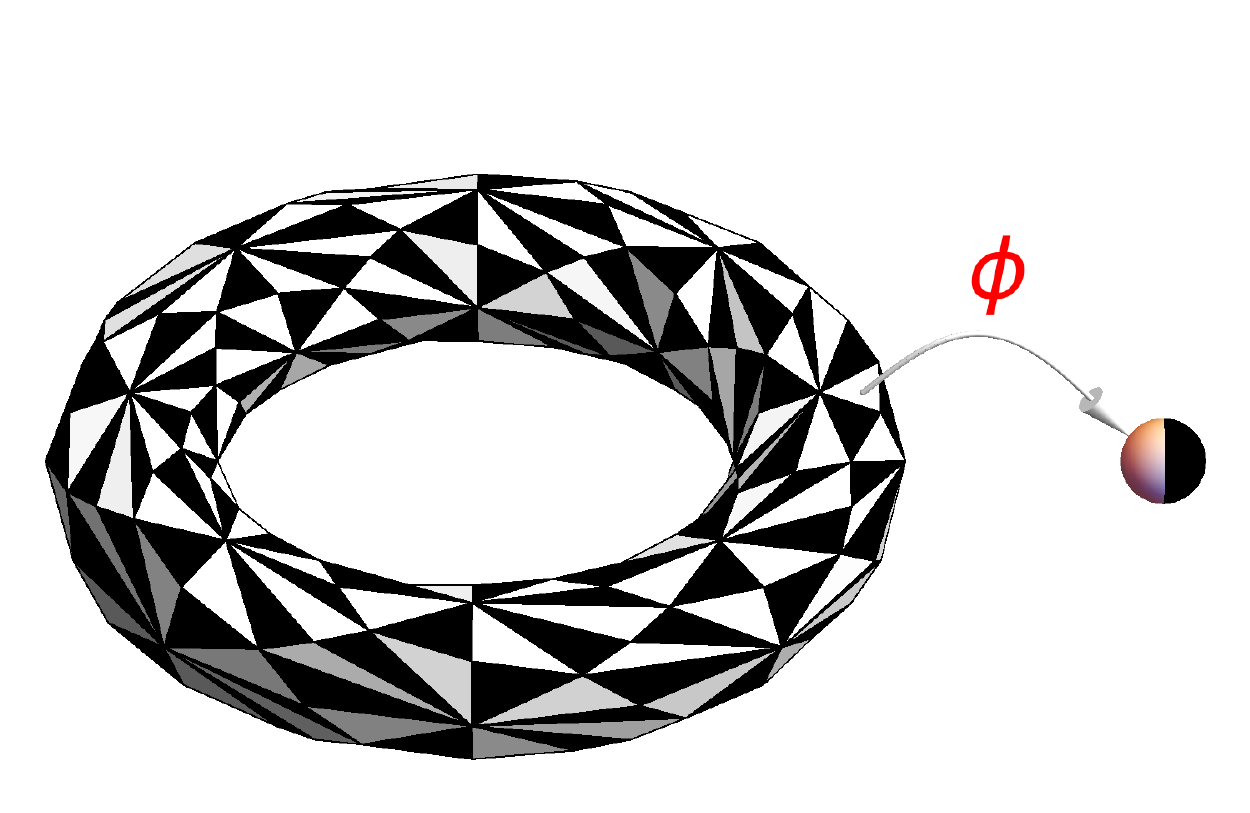}
%\includegraphics[scale=1]{triangul5.pdf}
%\end{center}
\caption{Triangulation of torus, the map $\phi$ maps white triangles to the
white hemisphere (of the small sphere) and the black ones to the black
hemisphere.}%
\label{torus}%
\end{figure}

\begin{figure}[ptb]
%\begin{center}
\includegraphics[scale=1]{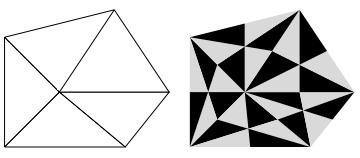}
%\end{center}
\caption{Barycentric subdivision.}%
\label{torus}%
\end{figure}

\subsection{Preliminaries in dimension $4$}

Let us first give simple examples in dimension $4$ of varieties where one can
obtain arbitrarily large quantized volumes.

First for the sphere $S^{4}$ itself one can construct by the same procedure as
in the proof of Lemma \ref{thm2} a smooth map $\phi:S^{4}\to S^{4}$ whose
Jacobian is $\geq0$ everywhere and whose degree is a given integer $N$. One
can then simply take the sum $\omega=\phi^{\#}(\alpha) +\alpha$ which does not
vanish and has integral $N+1$.

Next, let us take $M=S^{3}\times S^{1}$. Then one can construct by the same
procedure as in the proof of Lemma \ref{thm2} a smooth map $\phi:M\to S^{4}$
whose Jacobian is $\geq0$ everywhere and which vanishes only on a two
dimensional subset $K\subset M$. Let $p:M\to S^{3}$ be the first projection
using the product $M=S^{3}\times S^{1}$. Then $p(K)$ is a two dimensional
subset of $S^{3}$ and hence there exists $x\in S^{3}$, $x\notin p(K)$. One can
thus find a diffeomorphism $\psi\in\mathrm{Diff}^{+}(S^{3})$ such that
$\psi(p(K))\cap p(K)=\emptyset$. Then the diffeomorphism $\psi^{\prime}%
\in\mathrm{Diff}^{+}(M)$ which acts as $(x,y)\mapsto\psi^{\prime}(x,y)=
(\psi(x),y)$ is such that $\psi^{\prime}(K)\cap K=\emptyset$. Thus it follows
that the Jacobian of $\phi$ and the Jacobian of $\phi^{\prime}=\phi\circ
\psi^{\prime}$ never vanish simultaneously and the proof of Lemma \ref{thm2}
applies. Note moreover that in this case $M=S^{3}\times S^{1}$ is not simply
connected and one gets smooth covers of arbitrary degree which can be combined
with the maps $(\phi,\phi^{\prime})$.

\subsection{Necessary condition}

\label{sectnec}

Jean-Claude Sikorav and Bruno Sevennec found the following obstruction which
implies for instance that $D(\mathbb{C}P^{2})=\emptyset$. In general

\begin{lemma}
Let $M$ be an oriented compact smooth $4$-dimensional manifold, then, with
$w_{2}$ the second Stiefel-Whitney class of the tangent bundle,
\[
D(M)\neq\emptyset\implies w_{2}^{2}=0
\]
More generally if $D(M)\neq\emptyset$ and the dimension of $M$ is arbitrary,
the product of any two Stiefel-Whitney classes vanishes.
\end{lemma}

\proof
One has a cover of $M$ by two open sets on which the tangent bundle is stably
trivialized. Thus the product of any two Stiefel-Whitney classes vanishes.
\endproof

Since a manifold is a Spin manifold if and only if $w_{2}=0$ this obstruction
vanishes in our context.

\subsection{Reduction to a single map}

Here is a first lemma which reduces to properties of a single map.

\begin{lemma}
\label{single} Let $\phi:M\to S^{4}$ be a smooth map such that $\phi
^{\#}(\alpha)(x)\geq0$ $\forall x\in M$ and let $R=\{x\in M\mid\phi^{\#}
(\alpha)(x)=0\}$. Then there exists a map $\phi^{\prime}$ such that $\phi
^{\#}(\alpha)+\phi^{\prime\#}(\alpha)$ does not vanish anywhere if and only if
there exists an immersion $f:V\to\mathbb{R}^{4}$ of a neighborhood $V$ of $R$.
Moreover if this condition is fulfilled one can choose $\phi^{\prime}$ to be
of degree $0$.
\end{lemma}

\proof Let first $\phi^{\prime}$ be such that $\phi^{\#}(\alpha)+\phi
^{\prime\#}(\alpha)$ does not vanish anywhere. Then $\phi^{\prime\#}(\alpha)$
does not vanish on the closed set $R$ and hence in a neighborhood $V\supset
R$. Its restriction to $V$ gives the desired immersion. Conversely let
$f:V\to\mathbb{R}^{4}$ be an immersion of a neighborhood $V$ of $R$. We can
assume by changing the orientation of $\mathbb{R}^{4}$ for the various
connected components of $V$ that $f^{\#}(v)>0$ where $v$ is the standard
volume form on $\mathbb{R}^{4}$. We first extend $f$ to a smooth map $\tilde
f:M\to\mathbb{R}^{4}$ by extending the coordinate functions. We then can
assume that $f(M)\subset B_{4}\subset\mathbb{R}^{4}$ where $B_{4}$ is the unit
ball which we identify with the half sphere so that $B_{4}\subset S^{4}$. We
denote by $\beta=\alpha\vert B_{4}$ the restriction of $\alpha$ to $B_{4}$. We
have $f^{\#}(\beta)>0$ on $V$ but not on $M$ since the map $\tilde f:M \to
S^{4}$ is of degree zero. Let $\rho>0$ be a fixed volume form (nowhere
vanishing) on $M$. Let $\epsilon>0$ be such that
\[
\phi^{\#}(\alpha)(x)\geq\epsilon\rho(x), \ \ \forall x\notin V
\]
For $y\in B_{4}$ and $0<\lambda\leq1$ we let $\lambda y$ be the rescaled
element (using rescaling in $\mathbb{R}^{4}$). Then for $\lambda$ small enough
one has
\[
\vert(\lambda\tilde f)^{\#}(\alpha)(x)\vert\leq\frac12 \epsilon\rho(x),
\ \ \forall x\in M,
\]
where the absolute value is on the ratio of $(\lambda\tilde f)^{\#}(\alpha)$
with $\rho$. One then gets that with $\phi^{\prime}=\lambda\tilde f$ one has
\[
(\phi^{\#}(\alpha)+\phi^{\prime\#}(\alpha))(x)\neq0,\ \ \forall x\in M.
\]
\endproof

\subsection{Products $M=N\times S^{1}$}

Let $N$ be a smooth oriented compact three manifold. Then $N$ is Spin, thus
the condition $w_{2}^{2}=0$ is automatically fulfilled by $M=N\times S^{1}$.
In fact:

\begin{theorem}
\label{threeman}Let $N$ be a smooth oriented connected compact three manifold.
Let $M=N\times S^{1}$, then the set $q(M)$ is non-empty, and contains all
integers $m\geq r$ for some $r>0$.
\end{theorem}

\proof Let $g:S^{3}\times S^{1}\to S^{4}$ be a ramified cover of degree $m$
and singular set $\Sigma_{g}$. Let $N$ be described as a ramified cover
$f:N\to S^{3}$ ramified over a knot $K\subset S^{3}$ (\cite{Montesinos},
\cite{HM}). One may, using the two dimensionality of $\Sigma_{g}$, assume
that
\[
K\cap p_{3}(\Sigma_{g})=\emptyset, \ \ p_{3}: S^{3}\times S^{1}\to S^{3}.
\]
Let $h=f\times\mathrm{id}:N\times S^{1}\to S^{3}\times S^{1}$. Let $\Sigma
_{f}\subset N$ be the singular set of $f$. one has $f(\Sigma_{f})\subset K$
and thus, with $\Sigma_{h}\subset N\times S^{1}$ the singular set of $h$,
\[
\Sigma_{h}=\Sigma_{f}\times S^{1}, \ \ h(\Sigma_{h})\cap\Sigma_{g}=\emptyset
\]
since $h(\Sigma_{h})=f(\Sigma_{f})\times S^{1}\subset K\times S^{1}$ is
disjoint from $\Sigma_{g}$. Let then $\phi=g\circ h$. The singular set
$\Sigma_{\phi}$ of $\phi$ is the union of $\Sigma_{h}$ with $h^{-1}(\Sigma
_{g})$. This two closed sets are disjoint since $h(\Sigma_{h})\cap\Sigma
_{g}=\emptyset$. By Lemma \ref{single} it is enough to find immersions in
$\mathbb{R}^{4}$ of neighborhoods $V\supset\Sigma_{h}$ and $W\supset
h^{-1}(\Sigma_{g})$. By construction $\Sigma_{h}=\Sigma_{f}\times S^{1}$ is a
union of tori with trivial normal bundle, since their normal bundle is the
pullback by the projection of the normal bundle to $\Sigma_{f}$ which is a
union of circles. This gives the required immersion $V\to\mathbb{R}^{4}$.
Moreover the restriction of $h$ to a suitable neighborhood $W$ of
$h^{-1}(\Sigma_{g})$ is a smooth covering of an open set of $S^{3}\times
S^{1}$. On each of the components $W_{j}$ of this covering, the local
situation is the same as for the inclusion of $\Sigma_{g}$ in $S^{3}\times
S^{1}$. Thus one gets the required immersion $W\to\mathbb{R}^{4}$. This shows
that the hypothesis of Lemma \ref{single} is fulfilled and one gets that
$D(M)\neq\emptyset$ and that
\[
\mathrm{degree}(f)+\mathrm{degree}(g)\in q(M)
\]
\endproof

\begin{remark}
\textrm{Here is a variant, due to Jean-Claude Sikorav, of the above proof,
also using Lemma \ref{single}. The $4$-manifold $M=N\times S^{1}$ is
parallelizable since any oriented $3$-manifold is parallelizable (see for
instance \cite{HM} for a direct proof), and by \cite{Poenaru} Theorem 5, there
is an immersion $\psi:M\setminus\{p\}\to\mathbb{R}^{4}$ of the complement of a
single point $p\in M$ so that it is easy to verify the hypothesis of Lemma
\ref{single} and show that for any ramified cover $\phi:M\to S^{4}$ one has
$degree(\phi)\in q(M)$. }
\end{remark}

\subsection{Spin manifolds}

\begin{theorem}
\label{4man}Let $M$ be a smooth connected oriented compact spin $4$-manifold.
Then the set $q(M)$ contains all integers $m\geq5$.
\end{theorem}

\proof We proceed as in the proof of Lemma \ref{thm2} and get from any
Whitehead triangulation of $M$ (after a barycentric subdivision) a ramified
covering $\gamma$ of the sphere $S^{4}$ obtained by gluing two copies
$\Delta^{4}_{\pm}$ of the standard simplex $\Delta^{4}$ along their boundary.
Let then $V$ be a neighborhood of the $2$-skeleton of the triangulation which
retracts on the $2$-skeleton. Then the restriction of the tangent bundle of
$M$ to $V$ is trivial since the spin hypothesis allows one to view $TM$ as
induced from a $\mathrm{Spin}(4)$ principal bundle while the classifying space
$B\mathrm{Spin}(4)$ is $3$-connected. Thus the extension by Poenaru
\cite{Poenaru}, Theorem 5, (see also \cite{Phillips}), of the Hirsch-Smale
immersion theory (\cite{HS1}, \cite{HS2} \!\!) to the case of codimension zero
yields an immersion $V\to\mathbb{R}^{4}$. After smoothing $\gamma$ while
keeping its singular set inside $V$ one gets that the hypothesis of Lemma
\ref{single} is fulfilled and this gives that $m\in q(M)$ where $2m$ is the
number of simplices of the triangulation. For the finer result involving the
small values of $m$ one can use the theorem\footnote{This theorem is stated in
the PL category but, as confirmed to us by R. Piergallini, it holds (for any
$m\geq5$) in the smooth category due to general results PL=Smooth in
4-dimensions.} of M. Iori and R. Piergallini \cite{IP}, which gives a smooth
ramified cover $\phi:M\to S^{4}$ of any degree $m\geq5$ whose singular set
$R\subset M$ is a disjoint union of smooth surfaces $S_{j}\subset M$. As
above, when $M$ is a Spin manifold, the condition of Lemma \ref{single} is
fulfilled so that $m\in q(M)$. Indeed as above, this shows that there exists
an immersion of a neighborhood of each $S_{j}$ in $\mathbb{R}^{4}$. Thus
$q(M)$ contains any integer $m\geq5$ for any Spin $4$-manifold. \endproof

\begin{remark}
\textrm{In fact in the above proof one needs to use immersion theory only when
$S_{j}$ is non-orientable. If $S_{j}$ is orientable, then by Whitney's theorem
(\cite{WW0}, \S 6.b)) the Euler class $\chi(\nu)$ of the normal bundle of
$\phi(S_{j})\subset S^{4}$ is $\chi(\nu)=0$, while one has the proportionality
with the Euler class of the normal bundle $\nu^{\prime}$ of $S_{j}\subset M$.
Thus $\chi(\nu^{\prime})=0$ and it follows that there is an embedding of a
tubular neighborhood of $S_{j}$ in $\mathbb{R}^{4}$. }
\end{remark}

\begin{remark}
\textrm{As a countercheck it is important to note why the above proof does not
apply in the case of $\mathbb{C}P^{2}$ seen as a double cover of the
$4$-sphere which is the quotient of $\mathbb{C}P^{2}$ by complex conjugation
and gives a ramified cover with ramification on $\mathbb{R}P^{2}$. It is an
exercice for the reader to compute directly the second Stiefel-Whitney class
of the tangent space of $\mathbb{C}P^{2}$ restricted to the submanifold
$\mathbb{R}P^{2}$ and check that it does not vanish. }
\end{remark}

\begin{corollary}
\label{self} Let $M$ be a smooth compact connected oriented spin Riemannian
$4$-manifold with quantized\thinspace\thinspace\footnote{up to normalization}
volume $\geq5$. Then there exists an irreducible representation of the
two-sided quantization relation such that the canonical spectral triple
$(\mathcal{A},\mathcal{H},D)$ of $M$ appears as follows, where $\{Y^{A}%
,Y^{\prime B}\}^{\prime\prime}$ is the double commutant of the components
$Y^{A},Y^{\prime B}$,

$\bullet$~Algebra : $\mathcal{A}=\{f\in\{Y^{A},Y^{\prime B}\}^{\prime\prime
}\mid f\mathcal{D}\subset\mathcal{D}\}, \ \ \mathcal{D}= \cap_{k}%
\mathrm{Domain}D^{k}. $

$\bullet$~Hilbert space: $\mathcal{H}=\prod E_{A}E^{\prime}_{B} H$,
$E_{A}=\frac12 (1+\Gamma^{A})$, $E^{\prime}_{B}=\frac12 (1+\Gamma^{\prime B})$.

$\bullet$~Operator: The operator is the restriction of $D$ to $\mathcal{H}$.
\end{corollary}

\proof By Theorem \ref{4man} combined with Theorem \ref{bubles1}, a solution
of \eqref{jjj} exists for the spectral triple of $M$. Let $\phi,\phi^{\prime}$
be the corresponding maps $M\rightarrow S^{4}$. By a general position argument
(\cite{GG}, Chapter III, Corollary 3.3) one can assume that the map
$(\phi,\phi^{\prime}):M\to S^{4}\times S^{4}$ is transverse to itself, without spoiling
the fact that $\phi^{\#}(\alpha)+\phi^{\prime\#}(\alpha)$ does not vanish. The
existence of self-intersections of $M\subset S^{4}\times S^{4}$ prevents the
components $Y^{A},Y^{\prime B}$ from generating the algebra of smooth
functions on $M$ but what remains true is that the double commutant
$\{Y^{A},Y^{\prime B}\}^{\prime\prime}$ is the same as the double commutant of
$C^{\infty}(M)$ since the double points form a finite set. One then concludes
that, with $\mathcal{D}=\cap_{k}\mathrm{Domain}D^{k}$ one has
\[
C^{\infty}(M)=\{f\in\{Y^{A},Y^{\prime B}\}^{\prime\prime}\mid f\mathcal{D}%
\subset\mathcal{D}\}
\]
and it follows that the representation of the two-sided quantization relation
is irreducible. The formulas for the Hilbert space and the operator are
straightforward. \endproof

\section{A tentative particle picture in Quantum Gravity}

One of the basic conceptual ingredients of Quantum Field Theory is the notion
of particle which Wigner formulated as irreducible representations of the
Poincar\'e group. When dealing with general relativity we shall see that (in
the Euclidean $=$ imaginary time formulation) there is a natural corresponding
particle picture in which the irreducible representations of the two-sided
higher Heisenberg relation play the role of ``particles". Thus the role of the
Poincar\'e group is now played by the algebra of relations existing between
the line element and the slash of scalar fields.

We shall first explain why it is natural from the point of view of
differential geometry also, to consider the two sets of $\Gamma$-matrices and
then take the operators $Y$ and $Y^{\prime}$ as being the correct variables
for a first shot at a theory of quantum gravity. Once we have the $Y$ and
$Y^{\prime}$ we can use them and get a map $(Y,Y^{\prime}):M\to S^{n}\times
S^{n}$ from the manifold $M$ to the product of two $n$-spheres. The first
question which comes in this respect is if, given a compact $n$-dimensional
manifold $M$ one can find a map $(Y,Y^{\prime}): M\to S^{n}\times S^{n}$ which
embeds $M$ as a submanifold of $S^{n}\times S^{n}$. Fortunately this is a
known result, the strong embedding theorem of Whitney, \cite{WW}, which
asserts that any smooth real $n$-dimensional manifold (required also to be
Hausdorff and second-countable) can be smoothly embedded in the real 2n-space.
Of course ${\mathbb{R}}^{2n}={\mathbb{R}}^{n}\times{\mathbb{R}}^{n}\subset
S^{n}\times S^{n}$ so that one gets the required embedding. This result shows
that there is no restriction by viewing the pair $(Y,Y^{\prime})$ as the
correct \textquotedblleft coordinate" variables. Thus we simply view $Y$ and
$Y^{\prime}$ as operators in Hilbert space and we shall write algebraic
relations which they fulfill relative to the two Clifford algebras $C_{\kappa
}$, $\kappa=\pm1$ and to the self-adjoint operator $D$. We should also involve
the $J$ and the $\gamma$. The metric dimension will be governed by the growth
of the spectrum of $D$.

The next questions are: assuming that we now no-longer use a base manifold $M$,

\begin{enumerate}
\item[A:] Why is it true that the joint spectrum of the $Y^{A}$ and $Y^{\prime
B}$ is of dimension $n$ while one has $2n$ variables.

\item[B:] Why is it true that the non-commutative integrals
\[
{\int\!\!\!\!\!\! -} \gamma\left\langle Y\left[  D,Y\right]  ^{n}\right\rangle
D^{-n}, \ \ {\int\!\!\!\!\!\! -} \gamma\left\langle Y^{\prime}\left[
D,Y^{\prime}\right]  ^{n}\right\rangle D^{-n}, \ \ {\int\!\!\!\!\!\! -}D^{-n}
\]
remain quantized.
\end{enumerate}

\subsection{Why is the joint spectrum of dimension $4$}

The reason why $A$ holds in the case of classical manifolds is that in that
case the joint spectrum of the $Y^{A}$ and $Y^{\prime B}$ is the subset of
$S^{n}\times S^{n}$ which is the image of the manifold $M$ by the map $x\in
M\mapsto(Y(x),Y^{\prime}(x))$ and thus its dimension is at most $n$.

The reason why $A$ holds in general is because of the assumed boundedness of
the commutators $[D,Y]$ and $[D,Y^{\prime}]$ together with the commutativity
$[Y,Y^{\prime}]=0$ (order zero condition) and the fact that the spectrum of
$D$ grows like in dimension $n$.

\subsection{Why is the volume quantized}

The reason why $B$ holds in the case of classical manifolds is that this is a
winding number, as shown in Lemma \ref{bublecor}.

The reason why $B$ holds in the general case is that all the lower components
of the operator theoretic Chern character of the idempotent $e=\frac12 (1+Y)$
vanish and this allows one to apply the operator theoretic index formula which
in that case gives (up to suitable normalization)
\[
2^{-n/2-1}{\int\!\!\!\!\!\! -} \gamma\left\langle Y\left[  D,Y\right]
^{n}\right\rangle D^{-n}= \mathrm{Index}\, (D_{e})
\]
This follows from the local index formula of \cite{CM} but in fact one does
not need the technical hypothesis of \cite{CM} since, when the lower
components of the operator theoretic Chern character all vanish, one can use
the non-local index formula in cyclic cohomology and the determination in
\cite{Connesbook} Theorem 8, IV.2.$\gamma$ of the Hochschild class of the
index cyclic cocycle.

To be more precise one introduces the following trace operation, given an
algebra $\mathcal{A}$ over $\mathbb{R}$ (not assumed commutative) and the
algebra $M_{n}(\mathcal{A})$ of matrices of elements of $\mathcal{A}$, one
defines
\[
\mathrm{tr}:M_{n}(\mathcal{A})\otimes M_{n}(\mathcal{A})\otimes\cdots\otimes
M_{n}(\mathcal{A})\rightarrow\mathcal{A}\otimes\mathcal{A}\otimes\cdots
\otimes\mathcal{A}%
\]
by the rule, using $M_{n}(\mathcal{A})=M_{n}(\mathbb{R})\otimes\mathcal{A}$
\[
\mathrm{tr}\left(  (a_{0}\otimes\mu_{0})\otimes(a_{1}\otimes\mu_{1}%
)\otimes\cdots\otimes(a_{m}\otimes\mu_{m})\right)  =\mathrm{Trace}(\mu
_{0}\cdots\mu_{m})a_{0}\otimes a_{1}\otimes\cdots\otimes a_{m}%
\]
where $\mathrm{Trace}$ is the ordinary trace of matrices. Let us denote by
$\iota_{k}$ the operation which inserts a $1$ in a tensor at the $k$-th place.
So for instance
\[
\iota_{0}(a_{0}\otimes a_{1}\otimes\cdots\otimes a_{m})=1\otimes a_{0}\otimes
a_{1}\otimes\cdots\otimes a_{m}%
\]
One has $\mathrm{tr}\circ\iota_{k}=\iota_{k}\circ\mathrm{tr}$ since (taking
$k=0$)
\[
\mathrm{tr}\circ\iota_{0}\left(  (a_{0}\otimes\mu_{0})\otimes(a_{1}\otimes
\mu_{1})\otimes\cdots\otimes(a_{m}\otimes\mu_{m})\right)  =
\]%
\[
=\mathrm{tr}\left(  (1\otimes1)\otimes(a_{0}\otimes\mu_{0})\otimes
(a_{1}\otimes\mu_{1})\otimes\cdots\otimes(a_{m}\otimes\mu_{m})\right)
\]%
\[
=\mathrm{Trace}(1\mu_{0}\cdots\mu_{m})1\otimes a_{0}\otimes a_{1}\otimes
\cdots\otimes a_{m}=
\]%
\[
=\iota_{0}\left(  \mathrm{tr}\left(  (a_{0}\otimes\mu_{0})\otimes(a_{1}%
\otimes\mu_{1})\otimes\cdots\otimes(a_{m}\otimes\mu_{m})\right)  \right)
\]
The components of the Chern character of an idempotent $e\in M_{s}%
(\mathcal{A})$ are then given up to normalization by
\begin{equation}
\mathrm{Ch}_{m}(e):=\mathrm{tr}\left(  (2e-1)\otimes e\otimes e\otimes
\cdots\otimes e\right)  \in\mathcal{A}\otimes\mathcal{A}\otimes\ldots
\otimes\mathcal{A} \label{phim}%
\end{equation}
with $m$ even and equal to the number of terms $e$ in the right hand side. Now
the main point in our context is the following general fact

\begin{lemma}
\label{lowerchern} Let $\mathcal{A}$ be an algebra (over $\mathbb{R}$) and
$Y=\sum Y^{A}\Gamma_{A}$ with $Y^{A}\in\mathcal{A}$ and $\Gamma_{A}\in
C_{+}\subset M_{w}(\mathbb{C})$ as above, $n+1$ gamma matrices. Assume that
$Y^{2}=1$. Then for any even integer $m<n$ one has $\mathrm{Ch}_{m}(e)=0$
where $e=\frac{1}{2}(1+Y)$.
\end{lemma}

\proof This follows since the trace of a product of $m+1$ gamma matrices is
always $0$. \endproof

It follows that the component $\mathrm{Ch}_{n}(e)$ is a Hochschild cycle and
that for any cyclic $n$-cocycle $\phi_{n}$ the pairing $<\phi_{n},e>$ is the
same as $<I(\phi_{n}),\mathrm{Ch}_{n}(e)>$ where $I(\phi_{n})$ is the
Hochschild class of $\phi_{n}$. This applies to the cyclic $n$-cocycle
$\phi_{n}$ which is the Chern character $\phi_{n}$ in $K$-homology of the
spectral triple $(\mathcal{A},\mathcal{H},D)$ with grading $\gamma$ where
$\mathcal{A}$ is the algebra generated by the components $Y^{A}$ of $Y$ and
$Y^{\prime A}$ of $Y^{\prime}$. By \cite{Connesbook} Theorem 8, IV.2.$\gamma$,
(see also \cite{FGV} Theorem 10.32 and \cite{CRSZ} for recent optimal
results), the Hochschild class of $\phi_{n}$ is given, up to a normalization
factor, by the Hochschild $n$-cocycle:
\[
\tau(a_{0},a_{1},\ldots,a_{n})={\int\!\!\!\!\!\!-}\gamma a_{0}[D,a_{1}%
]\cdots\lbrack D,a_{n}]D^{-n},\ \ \forall a_{j}\in\mathcal{A}.
\]
Thus one gets that, by the index formula, for any idempotent $e\in M_{s}(A)$
\[
<\tau,\mathrm{Ch}_{n}(e)>=<\phi_{n},e>=\mathrm{Index}\,(D_{e})\in\mathbb{Z}%
\]
Now by \eqref{phim} for $m=n$ and the fact that $D$ commutes with the two
Clifford algebras $C_{\pm}$, one gets, with $Y=2e-1$ as above, the formula
\[
<\tau,\mathrm{Ch}_{n}(e)>={\int\!\!\!\!\!\!-}\gamma\left\langle Y\left[
D,Y\right]  ^{n}\right\rangle D^{-n}%
\]
The same applies to $Y^{\prime}$ and we get

\begin{theorem}
\label{thm3} The quantization equation implies that (up to normalization)
\[
{\int\!\!\!\!\!\! -} D^{-n}\in\mathbb{N }%
\]

\end{theorem}

\proof One has, from the two sided equation,
\[
\frac{1}{n!}\left\langle Z\left[  D,Z\right]  ^{n}\right\rangle =\gamma
\]
so that
\[
{\int\!\!\!\!\!\!-}D^{-n}={\int\!\!\!\!\!\!-}\gamma\gamma D^{-n}=\frac{1}%
{n!}{\int\!\!\!\!\!\!-}\gamma\left\langle Z\left[  D,Z\right]  ^{n}%
\right\rangle D^{-n}%
\]
and using \eqref{quant}
\[
{\int\!\!\!\!\!\!-}\gamma\left\langle Z\left[  D,Z\right]  ^{n}\right\rangle
D^{-n}=\frac12{\int\!\!\!\!\!\!-}\gamma\left\langle Y\left[  D,Y\right]
^{n}\right\rangle D^{-n}+\frac12{\int\!\!\!\!\!\!-}\gamma\left\langle
Y^{\prime}\left[  D,Y^{\prime}\right]  ^{n}\right\rangle D^{-n}%
\]
which gives the required result after a suitable choice of normalization since
both terms on the right hand side give indices of Fredholm
operators.\endproof

\section{Conclusions}

In this paper we have uncovered a higher analogue of the Heisenberg
commutation relation whose irreducible representations provide a tentative
picture for quanta of geometry. We have shown that $4$-dimensional Spin
geometries with quantized volume give such irreducible representations of the
two-sided relation involving the Dirac operator and the Feynman slash of
scalar fields and the two possibilities for the Clifford algebras which
provide the gamma matrices with which the scalar fields are contracted. These
instantonic fields provide maps $Y,Y^{\prime}$ from the four-dimensional
manifold $M_{4}$ to $S^{4}.$ The intuitive picture using the two maps from
$M_{4}$ to $S^{4}$ is that the four-manifold is built out of a very large
number of the two kinds of spheres of Planckian volume. The volume of
space-time is quantized in terms of the sum of the two winding numbers of the
two maps. More suggestively the Euclidean space-time history unfolds to
macroscopic dimension from the product of two $4$-spheres of Planckian volume
as a butterfly unfolds from its chrysalis. Moreover, amazingly, in dimension
$4$ the algebras of Clifford valued functions which appear naturally from the
Feynman slash of scalar fields coincide exactly with the algebras that were
singled out in our algebraic understanding of the standard model using
noncommutative geometry thus yielding the natural guess that the spectral
action will give the unification of gravity with the Standard Model (more
precisely of its asymptotically free extension as a Pati-Salam model as
explained in \cite{CCS}).

Having established the mathematical foundation for the quantization of
geometry, we shall present consequences and physical applications of these
results in a forthcoming publication \cite{CCM}.

\section*{Acknowledgement}

AHC is supported in part by the National Science Foundation under Grant No.
Phys-1202671. The work of VM is supported by TRR 33 \textquotedblleft The Dark
Universe\textquotedblright\ and the Cluster of Excellence EXC 153
\textquotedblleft Origin and Structure of the Universe\textquotedblright. AC
is grateful to Simon Donaldson for his help on Lemma \ref{single}, to Blaine
Lawson for a crash course on the $h$-principle and to Thomas Banchoff for a
helpful discussion. The solution of the problem of determining exactly for
which $4$-manifolds one has $D(M)\neq\emptyset$ has been completed by Bruno
Sevennec in \cite{SS} where he shows that it is equivalent to the necessary
condition found by Jean-Claude Sikorav and himself, namely $w_{2}^{2}=0$ as
described above in \S \ref{sectnec}.


\begin{thebibliography}{99}                                                                                               %


\bibitem {Alex}J. Alexander, \textit{Note on Riemann spaces}. Bull. Amer.
Math. Soc. 26 (1920), no. 8, 370-372.

\bibitem {CRSZ}A.L. Carey, A. Rennie, F. Sukochev, D. Zanin, \textit{
Universal measurability and the Hochschild class of the Chern character}
arXiv: 1401.1860.

\bibitem {CC}A. H. Chamseddine and A. Connes, \textit{The uncanny precision of
the spectral action} , Comm. Math. Phys. \textbf{293}, 867-897.

\bibitem {CC2}A. H. Chamseddine and A. Connes, \textit{Why the Standard Model}
, J. Geom. Phys. \textbf{58}, (2008) 38-47.

\bibitem {CCS}A. H. Chamseddine, A. Connes, W. van Suijlekom, \textit{Beyond
the Spectral Standard Model: Emergence of Pati-Salam Unification} , JHEP
\textbf{11 }(2013) 132.

\bibitem {CCM}A. H. Chamseddine, Alain Connes and V. Mukhanov, in preparation.

\bibitem {Connesbook}A. Connes, \textit{Noncommutative Geometry}. Academic
Press, 1994.

\bibitem {Connes}A. Connes, \textit{A short Survey of Noncommutative
Geometry}, J. Math. Phys. 41 (2000), no. 6, 3832-3866.

\bibitem {CM}A. Connes, H. Moscovici \textit{The local index formula in
noncommutative geometry}. Geom. Funct. Anal. 5 (1995), no. 2, 174-243.

\bibitem {GG}M. Golubitsky, V. Guillemin, \textit{Stable mappings and their
singularities}. Graduate Texts in Mathematics, Vol. 14. Springer-Verlag, New
York-Heidelberg, 1973.

\bibitem {FGV}J. Gracia-Bondia, J. Varilly, H. Figueroa, \textit{Elements of
noncommutative geometry}. Birkhauser Advanced Texts: Basler Lehrbacher.
[Birkhauser Advanced Texts: Basel Textbooks] Birkhauser Boston, Inc., Boston,
MA, 2001.

\bibitem {Greub}W. Greub, S. Halperin and R. Vanstone, \textit{Connections,
Curvature and Cohomology, }volumes 1-3, and in particular pages 347-351 volume
2 (sphere maps).

\bibitem {HM}H. Hilden, J. Montesinos, T. Thickstun, \textit{Closed oriented
3-manifolds as 3-fold branched coverings of $S^{3}$ of special type}. Pacific
J. Math. 65 (1976), no. 1, 65-76.

\bibitem {HS1}M. Hirsch, \textit{ Immersions of manifolds}. Transactions
A.M.S. 93 (1959), 242-276.

\bibitem {IP}M. Iori, R. Piergallini, \textit{$4$-manifolds as covers of the
$4$-sphere branched over non-singular surfaces}. Geometry and Topology, 6
(2002) 393-401.

\bibitem {Montesinos}J. Montesinos, \textit{A representation of closed
orientable 3-manifolds as 3-fold branched coverings of $S^{3}$}. Bull. Amer.
Math. Soc. 80 (1974), 845-846.

\bibitem {Moser}J. Moser, \textit{ On the volume elements on a manifold}.
Trans. Amer. Math. Soc. 120 1965 286-294.

\bibitem {Poenaru}V. Poenaru, \textit{Sur\ la\ th\'{e}orie\ des\ immersions}.
Topology 1 (1962) 81-100.

\bibitem {Phillips}A. Phillips, \textit{Submersions of open manifolds}.
Topology 6 (1967) 171-206.

\bibitem {Ramirez}A. Ramirez, \textit{Sobre un teorema de Alexander} An. Inst.
Mat. Univ. Nac. Autonoma Mexico, 15, No 1 (1975) 77-81.

\bibitem {SS}B. Sevennec, \textit{Janus immersions in the product of two
n-spheres}.

\bibitem {HS2}S. Smale, \textit{The classification of immersions of spheres in
Euclidean spaces}. Ann. Math. 69 (1959), 327-344

\bibitem {WW0}H. Whitney, \textit{On the topology of differentiable
manifolds}. Lectures in Topology, pp. 101-141. University of Michigan Press,
Ann Arbor, Mich., 1941.

\bibitem {WW}H. Whitney, \textit{The self-intersection of a smooth
$n$-manifold in $2n$-space}, Annals of Math. 45 Vol 2 (1944) 220-246.
\end{thebibliography}
\end{document}